\documentclass[conference, 10pt]{IEEEtran}
\IEEEoverridecommandlockouts 
\usepackage[]{graphicx}
\graphicspath{{Figures/}}
\usepackage{caption}
\usepackage{subfig} 
\usepackage{amsmath}
\usepackage{amssymb}
\usepackage{epsfig}
\usepackage{cite}
\usepackage{color}
\usepackage{balance}
\usepackage{amsmath}
\usepackage{amsfonts} 
\usepackage{graphicx}
\usepackage{pdfpages}
\usepackage[T1]{fontenc} 
\usepackage{array}
\usepackage{url}

\usepackage{color}
\usepackage{wrapfig}
\usepackage{lipsum}
\usepackage[hidelinks]{hyperref}
\usepackage{multirow}
\usepackage{tabularx}
\usepackage{tikz}
\usepackage{nth} 
\usepackage{algpseudocode} 
\usepackage{algorithm,algpseudocode}
\usepackage{breqn}
\begin{document}
\title{On the Importance of PWV in \\Detecting Precipitation}
\author{\IEEEauthorblockN{Shilpa Manandhar\IEEEauthorrefmark{1},
Soumyabrata Dev\IEEEauthorrefmark{2},
Yee Hui Lee\IEEEauthorrefmark{1} and
Yu Song Meng\IEEEauthorrefmark{3}}
\IEEEauthorblockA{\IEEEauthorrefmark{1}School of Electrical and Electronic Engineering, Nanyang Technological University (NTU), Singapore}
\IEEEauthorblockA{\IEEEauthorrefmark{2}The ADAPT Centre, School of Computer Science and Statistics, Trinity College Dublin, Ireland}
\IEEEauthorblockA{\IEEEauthorrefmark{3}National Metrology Centre, Agency for Science, Technology and Research (A$^{*}$STAR), Singapore}
\thanks{This research is funded by the Defence Science and Technology Agency (DSTA), Singapore.}
\thanks{Send correspondence to Y.\ H.\ Lee, E-mail: EYHLee@ntu.edu.sg.}
\vspace{-0.6cm}
}

\maketitle

\begin{abstract}
With a rapid increase in the number of geostationary satellites around the earth's orbit, there has been a renewed interest in using Global Positioning System (GPS) to understand several phenomenon in earth's atmosphere. Such study using GPS devices are popular amongst the remote sensing community, as they provide several advantages with respect to scalability and range of applications. In this paper, we discuss how GPS signals can be used to estimate the amount of water vapor in the atmosphere. Furthermore, we demonstrate the importance of such precipitable water vapor (PWV) in the atmosphere for the task of rainfall detection. We present a detailed analysis in our dataset of meteorological data of $3$ years. Test dataset shows that use of PWV in rainfall detection helps to reduce the false alarm rate by almost 12\%.
\end{abstract}

\IEEEpeerreviewmaketitle

\section{Introduction}
Global positioning system satellites are now-a-days extensively used to study the total water content in the atmosphere (also known as precipitable water vapor) and surface air temperature~\cite{fujita2017observed}. This is important to study as PWV impacts the hydrological cycle and other weather phenomenon. In this paper~\footnote{The source code of all simulations in this paper is available online at \url{https://github.com/shilpa-manandhar/PWV-for-rainfall}}, we study the importance of PWV for the task of rainfall detection. Using a set of weather- and temporal- parameters, we use a data-centric methodology to establish this fact. 

\section{GPS \& Precipitation}
The amount of precipitable water vapor in the atmosphere can be estimated from the GPS signal delays~\cite{shilpaPI}. The delay incurred by the GPS signals is often referred as the zenith wet delay (ZWD). We calculate the PWV values (measured in mm), from the ZWD delays via:

\begin{equation}
\mbox{PWV}=\frac{PI \cdot \delta L_w^{o}}{\rho_l},
\end{equation} 

where \textit{$\delta$L$_w^{o}$} is ZWD, $\rho_{l}$ is the density of liquid water (1000 kg$/m^{3}$), and $PI$ is a dimensionless constant. We compute this parameter $PI$ using:

\begin{dmath}
	PI=[-1\cdot sgn(L_{a})\cdot 1.7\cdot 10^{-5} |L_{a}|^{h_{fac}}-0.0001]\cdot cos(\frac{(DoY-28)2\pi}{365.25})+[0.165-(1.7\cdot 10^{-5})|L_{a}|^{1.65}]+f.
\end{dmath}

In this equation, \textit{L$_{a}$} is the latitude, \textit{DoY} is day-of-year, \textit{h$_{fac}$} is $1.48$ for stations from northern- and $1.25$ for stations from southern- hemisphere. The $f$ is computed using $f=-2.38\cdot 10^{-6}H$, where \textit{H} is the station height. We compute the PWV values for the IGS GPS station (station ID: NTUS), situated at Nanyang Technological University (NTU). These PWV values have a resolution of $5$ minutes and have shown to be a good indicator of rainfall \cite{IGARSS_2016}

In our experiments, the meteorological parameters recorded by Davis Instruments 7440 Weather Vantage Pro II at a particular rooftop (1.3$^{\circ}$N, 103.68$^{\circ}$E) of NTU building are used. Different parameters like temperature, relative humidity, solar irradiance, wind speed, direction and rainfall rate are recorded. All these weather measurements are recorded at an interval of $1$ minute. In addition to these weather sensor data, we also collect continuous stream of sky images using collocated whole sky imagers. These images captured by our sky camera~\cite{WAHRSIS,IGARSS2015a} provide a visual understanding of the various phenomenon in the earth's atmosphere. In our previous work~\cite{rainonset}, we used such sky camera images to detect the onset of precipitation. 

\section{Rainfall Detection}

\begin{table*}[htb]
\normalsize
\centering
\caption{Performance of rainfall detection for our SVM-based model. All values are indicated in percentage (\%). For the case where PWV is considered, an increase in performance is indicated by $\uparrow$, and a decrease in performance is indicated by $\downarrow$.}
\label{tab:pwv_results}
\begin{tabular}{|p{3cm}|c|c|c|c|}
\hline
\multicolumn{1}{|c|}{\multirow{2}{*}{}} & \multicolumn{2}{c|}{Without PWV feature} & \multicolumn{2}{c|}{With PWV feature} \\ \cline{2-5} 
\multicolumn{1}{|c|}{}                  & True Detection (TD)      & False Alarm (FA)      & True Detection (TD)    & False Alarm (FA)    \\ \hline
Testing data                              & 92.6                 & 44.0              & 87.4 ($\downarrow$)   & 32.2 ($\uparrow$)          \\ \hline
Validation data                       & 88.2                 & 33.6              & 85.8 ($\downarrow$)               & 28.5 ($\uparrow$)           \\ \hline
\end{tabular}
\end{table*}

The task of rainfall detection can be modeled as a supervised learning task, based on a set of features. We use a combination of weather- and temporal- variables as discriminatory features of this task. We consider $7$ parameters as features -- temperature, dewpoint temperature, relative humidity, solar radiation, PWV, time of day and day of year. These variables are used as features for training a Support Vector Machine (SVM) model in the detection of rainfall.

In our experiments, we consider all the weather observations for the year of $2010$ and $2011$ as the data observations for our SVM model. An independent year of $2012$ is used for the validation set. However, this datasets are highly unbalanced as the number of \emph{rain} observations is low, as compared to \emph{no rain} observations. Therefore, we employ random down sampling technique to make the dataset balanced. We employ a $1$:$1$ ratio for \emph{rain} to \emph{no rain} observations. In order to reduce random sample bias, we perform our experiments for randomly sampled $100$ trials. The average result of these $100$ trials are reported. 

For an objective evaluation in understanding the importance of PWV, we compute two metrics -- True Detection (TD) percentage and False Alarm (FA) percentage. The metrics TD and FA are defined as:

\begin{equation}
TD = \frac{TP}{TP+FN}
\end{equation}

\begin{equation}
FA = \frac{FP}{TN+FP},
\end{equation}

where $TP$, $TN$, $FP$ and $FN$ are true positive, true negative, false positive and false negative respectively, with respect to rainfall detection task. We consider $20$\% of the data in the balanced dataset of $2010$ and $2011$, as the training set. The remaining observations are considered as the testing set. We perform two experiments to prove the importance of PWV in detecting rainfall. We consider all the $7$ considered features, and compute the TD- and FA- percentage. The results are evaluated for testing- and validation- dataset. Furthermore, we consider a $6$ dimensional feature vector without considering the parameter PWV. 

The results are reported in Table~\ref{tab:pwv_results}. We observe that, if we consider PWV in our model, the false alarm performance greatly improves, as compared to the case where PWV is not considered. The FA decreases by $11.8\%$ for testing data, and the decreases by $5.1\%$ for the validation set. However, there is a slight degradation in the performance of true detection percentage. The TD slightly decreases by $5.2\%$ and $2.4\%$ for the testing- and validation- data respectively. These experiments are conducted via down-sampling techniques at $20\%$ training dataset.

\begin{figure}[h!]
\begin{center}
\includegraphics[width=0.45\textwidth]{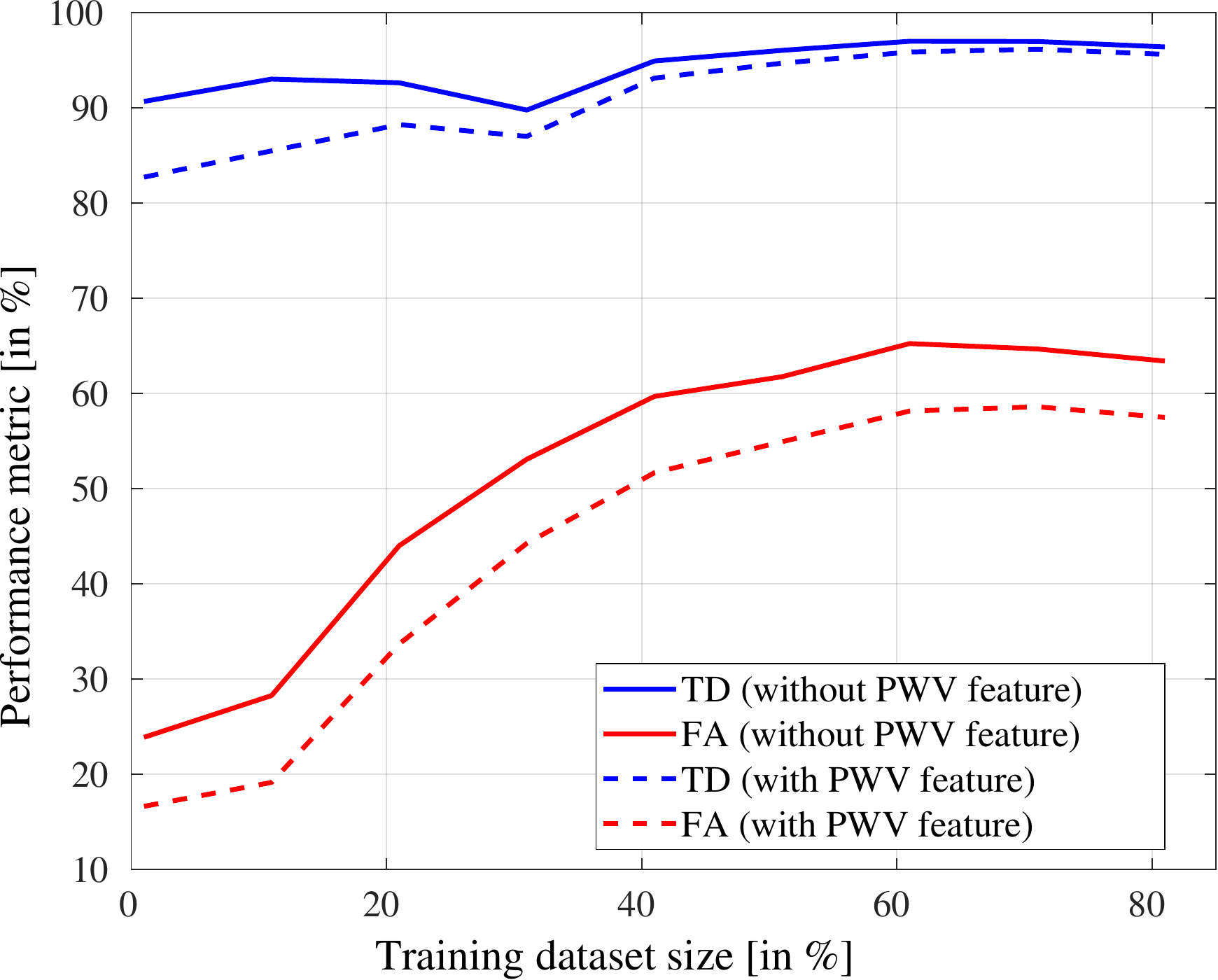}
\caption{True detection- and false alarm- percentage of rainfall recognition for varying percentage of training dataset.
\label{fig:diff-percent}}
\end{center}
\vspace{-0.5cm}
\end{figure}

We also perform experiments with varying percentage of training dataset. Figure~\ref{fig:diff-percent} shows the TD and FA percentages for varying percentage of training dataset. All these observations are the average of $100$ trials, that are randomly chosen from the dataset. It is clear that the performance of false alarm is significantly better, when PWV is considered in the model, for all the varying training size. However, the performance of the true detection slightly reduces by considering PWV. for smaller training dataset size. This difference in performance of TD diminishes for large training dataset size. Furthermore, we observe an over-saturation effect after $60\%$ of training size -- the TD and FA performances remains the same. This is because no further discriminatory variability of the dataset gets incorporated in our model, with the increasing training size.

\section{Conclusion \& Future Work}
The detection of water vapor content in the atmosphere is important amongst the remote sensing community to understand several phenomenon in the earth's atmosphere. It is especially, important for the detection and prediction of rainfall. In this paper, we have demonstrated the importance of PWV for the task of precipitation detection. Using a machine-learning framework, we observed that the false alarm reduces, when PWV is considered as one of the discriminatory features. In the future, we also plan to integrate image-based features extracted from ground-based sky cameras~\cite{Dev2016GRSM} in a multi-modal fashion, for further improving the current benchmark. 

\balance

\end{document}